# Monte Carlo Based Toy Model for Fission Process

R. Kurniadi, A. Waris, S. Viridi

Nuclear Physics and Biophysics Research Division

*Abstract*

Fission yield has been calculated notoriously by two calculations approach, macroscopic approach and microscopic approach. This work will proposes another calculation approach which the nucleus is treated as a toy model. The toy model of fission yield is a preliminary method that use random number as a backbone of the calculation. Because of nucleus as a toy model hence the fission process does not represent real fission process in nature completely. Fission event is modeled by one random number. The number is assumed as width of distribution probability of nucleon position in compound nuclei when fission process is started. The toy model is formed by Gaussian distribution of random number that randomizes distance like between particle and central point. The scission process is started by smashing compound nucleus central point into two parts that are left central and right central points. These three points have different Gaussian distribution parameters such as mean ($\mu_{CN}$, $\mu_L$, $\mu_R$), and standard deviation ($\sigma_{CN}$, $\sigma_L$, $\sigma_R$). By overlaying of three distributions, the number of particles ($N_L$, $N_R$) that they are trapped by central points is obtained. This process is iterated until ($N_L$, $N_R$) become constant numbers. The yield is determined from portion of nuclei distribution which is proportional with portion of mass numbers. Smashing process is repeated by changing $\sigma_L$ and $\sigma_R$ randomly

**Introduction**

Knowledge on nuclear fission is needed for large range of applications such as production of neutron-rich radioactive beams, neutron spallation sources, nucleosynthesis [1], design of Gen IV [2] and accelerator-driven systems [3-6]. There are several systematic of fission yields such as Wahl's systematic [7], Moriyama-Ohnishi [8], and Katakura [9]. These methods are developed by semi empirical technique, the fission yield curves (FYC) are modeled by several Gaussian functions. Single peak FYC is modeled by one Gaussian function, double peak use two Gaussian function. The parameters of function such as mean, standard deviation are fitted with data. This method has a weakness that is strongly depending on data. Without the data, FYC are not produced. Another approach that they do not use semi empirical are micro and macro approach. Fission yield has been calculated notoriously by these approaches that are macroscopic approach [10-11] and microscopic approach [12-15].

Microscopic approach treats fission process at low energy as adiabatic system. Collective motion of the system can be described in terms of few collective variables characterizing the shape evolution of nucleus that is mass axial deformation. This parameter is needed to calculate flux that went through scission line. Fragment mass distribution is dependent upon this flux.

Macroscopic approach uses Brossa model [16] as base model, fragment mass distribution is calculated by Gaussian distribution where rupture neck position is determined by random process. This model is called Random Neck Rupture Model (RNRM). RNRM is widely used to obtain fragment mass distribution.

There is another microscopic approach which it uses random number. This method is called Fission Toy Model (FTM). As a toy, the model does not represent real fission process in nature. Base on toy model, it can be expected there is an analogy between the model and real fission process in nature. Fission Toy Model (FTM) is a model that uses Monte Carlo as basic technique to obtain fission yield curve. Although FTM uses Monte Carlo but then FTM is not real fission process as well as FREYA (Fission Yield Event Yield Algorithm) [17] and prompt gamma emission [18]. FTM is preliminary study about application of Monte Carlo to calculation of mass distribution from fission process. There is no microscopic interaction between nucleons; the interaction is simulated by distribution of nucleons in nucleus. The FTM treats nucleus as a toy that has A marbles, which the interaction among marbles is neglected. In the FTM, marbles are influenced by central force. Because of this force, marbles are distributed. In order to FTM is simply model, in this paper the Gaussian distribution is adopted to illustrate marbles distribution where there is no connection between FTM and Fermi gas model of nucleus [19]. FTM assumes marbles as nucleons that they have attribute as ideal gases. Standard deviation of marble distribution is pictured by another distribution curve that it called Standard Deviation Force Distribution – (SDFD). The SDFD must not have Gaussian form. The frequency in SDFD represents numbers of fission events. Each event is correspondingly with one standard deviation force (SDF) value. Particles are distributed by Gaussian distribution which SDF value as standard deviation of this particle distribution.

**The Model**

After projectile hit the toy nucleus, the central force is smashed into two forces. These forces are called left center force (LCF) and right center force (RCF). LCF and RCF avoid each other by relative velocity *v*. LCF and RCF are illustrated by two Gaussian particle distribution with $\mu_L, \mu_R, \sigma_L, \sigma_R$ as parameters. These are mean and deviation standard of LCF and RCF respectively. The parameters have value randomly hence intersection point of two Gaussian distribution also in random situation. Illustration about distribution function overlapping is shown below,

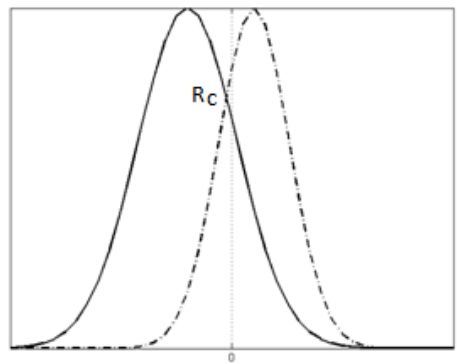

Figure 1. Distribution function overlapping in toy fission process

Figure 1 shows that two central forces avoid each other. LCF moves left and RCF moves right. At the $R_c - \epsilon$ point, marbles in this region interact with LCF; conversely marbles will interact with RCF if they are in $r \geq R_c + \epsilon$ region. Intersection point $R_c$ can be calculated by

$$R_c = \frac{\left(\mu_R/\sigma_R^2 - \mu_L/\sigma_L^2\right) + \sqrt{\left(\mu_R/\sigma_R^2 - \mu_L/\sigma_L^2\right)^2 - 2\left(1/\sigma_L^2 - 1/\sigma_R^2\right)\left(0.5\left(\frac{\mu_L}{\sigma_L}\right)^2 - 0.5\left(\frac{\mu_R}{\sigma_R}\right)^2 - \log\left(\frac{\sigma_R}{\sigma_L}\right)\right)}}{\left(1/\sigma_L^2 - 1/\sigma_R^2\right)} \quad (1)$$

In order that agree with normalization condition, it needs relation

$$\int_{-\infty}^{R_c} f_{LCF}(r)dr + \int_{R_c}^{\infty} f_{LCF}(r)dr = 1 \quad (2)$$

If equation (2) is multiplied by number of marbles A, the composition of marbles distribution is known that is

$$A \int_{-\infty}^{R_c} f_{LCF}(r)dr = N_L \quad (3)$$

$$A \int_{R_c}^{\infty} f_{LCF}(r)dr = N_R \quad (4)$$

$N_L$ and $N_R$ are number of marbles in left part and right part respectively. Explicitly, $N_L$ and $N_R$ are dependent upon position of center of LCF and RCF. At maximum elongation, composition of marbles distribution moves to constant values. Iterative formula of this composition has form

$$X_{k+1} = G X_k \quad (5)$$

where

$$G = \begin{pmatrix} 1 - G_{21} & 1 - G_{22} \\ 0.5 \, erf\left(\frac{R_c - \mu_R}{\sqrt{2}\sigma_R}\right) + 0.5 & 0.5 \, erf\left(\frac{R_c - \mu_L}{\sqrt{2}\sigma_L}\right) + 0.5 \end{pmatrix} \quad (6)$$

$$X_k = \begin{pmatrix} N_R \\ N_L \end{pmatrix} \quad (7)$$

Simplest model is constant value of $\mu_L$, $\mu_R$. In this work this parameters is calculated by formula

$$\mu_L = -\gamma R \alpha \tag{8a}$$

$$\mu_R = \gamma R \alpha \tag{8b}$$

$R$ is nucleus toy radius $R = r.A^{1/3}$ where $A$ is nucleus toy mass number. $\gamma$, $\alpha$ are coefficient and random number respectively. Each fission event is symbolized by $\alpha$ number.

The work flow diagram of FTM is showed by figure 2 below,

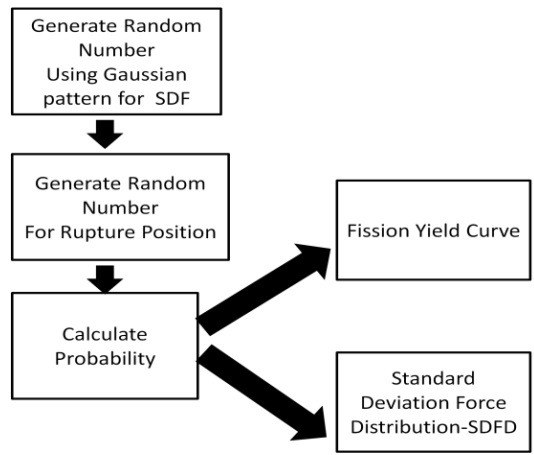

Figure 2. Work flow diagram of FTM

It is useful if SDFD had been created before FYC calculation is started. This algorithm is suitable if there is correlation between energy of real fission event and SDFD. The modification is shown by figure 3 below

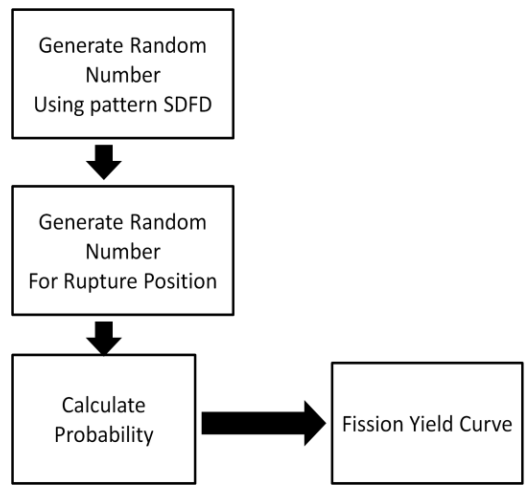

Figure 3. Work flow diagram of FTM modification

**Results**

This work will illustrates the fission process of A=238 nuclide toy. Neutron toy at energy $E_o$ hits the nuclide, which divides into two lighter nuclides. Each fission process is initiated by constructing central forces that are LC force and RC force. These forces are figured by distribution of marbles. The force strength is illustrated by $\sigma$, smaller $\sigma$ value gives higher force strength. $\sigma$ value does not have formulation directly to force strength. Figure 4 shows illustration of $\sigma$ distribution.

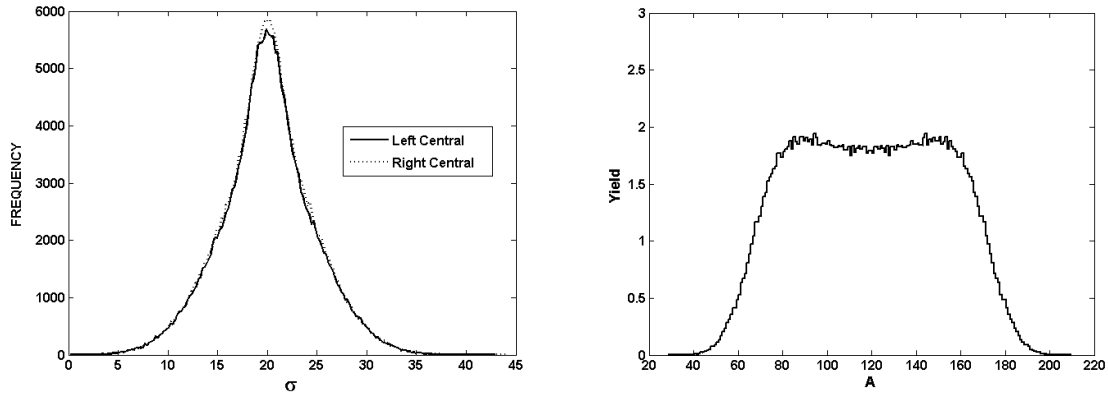

Figure 4. $\sigma$ Distribution of LC force and RC force (SDFD) (left) and Fission yield curve (right) .

This distribution is obtained from 2.4e+05 fission events. $\sigma$ gives information about marbles distribution around the RC or LC position. Highest $\sigma$ population is about 20 units, it has meaning that the marbles are distributed dominantly around central point with deviation is 20 units. Base on figure 1 (there is highly domination of single $\sigma$ value), It can be said that the fission is not focused at asymmetrical fission yield, the left and the right $\sigma$ distributions are almost identical. This distribution gives fission yield curve as figure 4 showed above.

In order to clarity of illustration is obtained, figure 5 (left) displayed another situation. Fission process is dominated by asymmetrical fission yield.

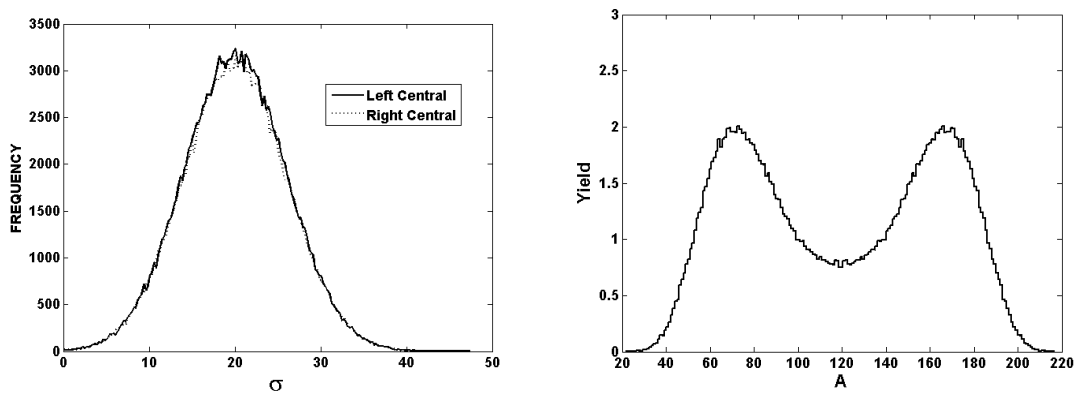

Figure 5. Another $\sigma$ distribution of LC force and RC force (SDFD) (left) and two peaks fission yield curve (right) .

The $\sigma$ distribution does not have symmetrical shape. In $\sigma > 20$ region the curve more steep than left region. Figure 5 (left) is generated by 2e+05 fission events, in these events there are variation of $\sigma$ values which left $\sigma$ and right $\sigma$ have different values in same event. The different value in same event gives different distribution of marbles hence asymmetrical $\sigma$ distribution causes fission yield curve has two peaks. If in figure 5 (left) $\sigma$ distribution is added another distribution curve that it has more incisive, the fission yield curve will has three peaks. This condition is shown figure 6 below,

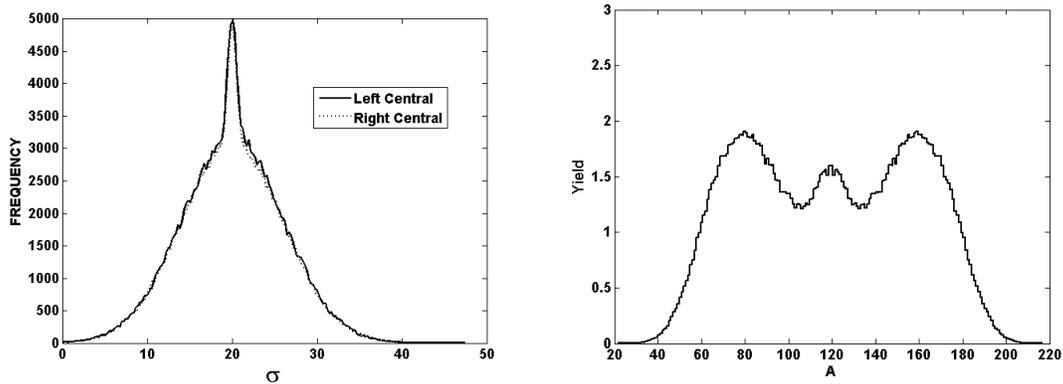

Figure 6. Another $\sigma$ distribution of LC force and RC force (left) and three peaks fission yield curve (right)

Incisive distribution gives narrowly marble opportunity to make more variation of $\sigma$ value hence fission yield is dominated by symmetrical result. Another modification is made by adding combination between two $\sigma$ distributions. For example, figure 7 (left) shows combination two different distributions.

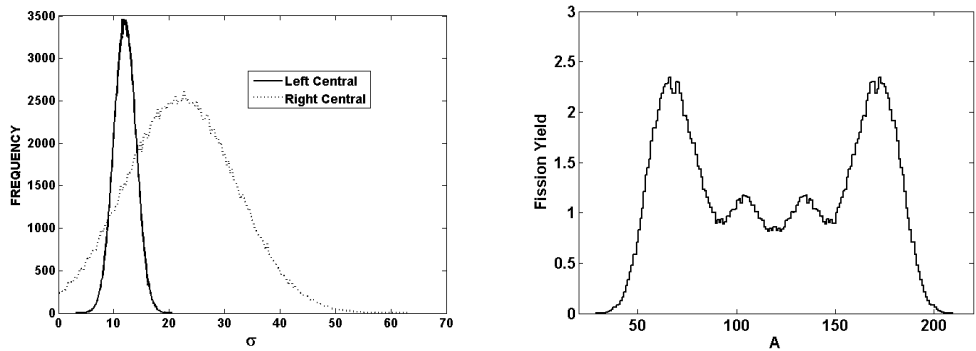

Figure 7. Another $\sigma$ distribution of LC force and RC force (left) and four peaks fission yield curve (right).

Figure 7 (left) demonstrates the two different shapes of $\sigma$ distributions, there are two differentiations between LC and RC that are mean ($\mu_\sigma$) and standard deviation ($std_\sigma$) of $\sigma$. By smaller $\mu_\sigma$ value, marbles are distributed about central whereas smaller $std_\sigma$ cause's marbles are trapped in short distribution. The LC curve in figure 7 (left) does not have same shape with figure 4 (left). Although they have sharp shape in $\mu_\sigma$ position, LC $\sigma$ distribution from figure 7 (left) has narrow curve at bottom frequency. By this condition, asymmetrical fission product emerges. The RC $\sigma$ distribution in figure 7 (left) has similar curve pattern with figure 5 (left). Both of distributions gives two peaks approximately around A = 70 and A = 170.

Figure 8 shows the convergence of the FTM. Figure 8 (left) and figure 8 (right) are obtained by $2.10^3$ iterations and $5.10^4$ iterations respectively.

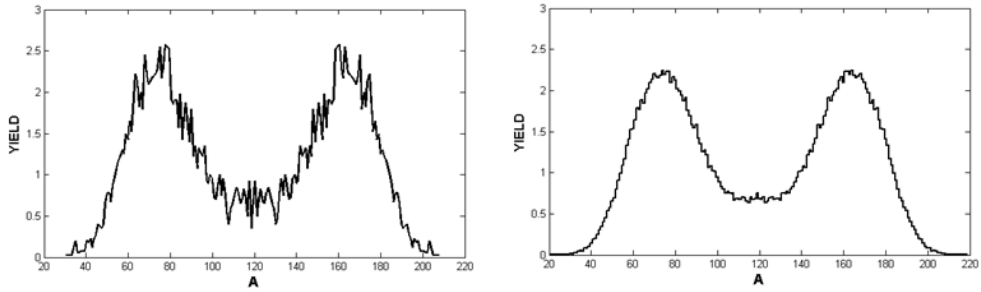

Figure 8. Fission yields are obtained by $2.10^3$ iterations (left) and $5.10^4$ iterations (right).

This figure shows clearly that more iteration will give better results. The $\gamma$ values in equation 8 strongly influences on $R_C$, then smaller $\gamma$ values gives smaller range of $\mu$, through this condition equation (1) is changed into

$$R_c = \frac{\sqrt{-2\left(1/\sigma_L^2 - 1/\sigma_R^2\right)\left(0.5\left(\frac{\mu_L}{\sigma_L}\right)^2 - 0.5\left(\frac{\mu_R}{\sigma_R}\right)^2 - \log\left(\frac{\sigma_R}{\sigma_L}\right)\right)}}{\left(1/\sigma_L^2 - 1/\sigma_R^2\right)} \qquad (9)$$

If $\gamma$ values very small then equation (9) becomes

$$R_c = \sqrt{2\left(\frac{\sigma_L^2 \sigma_R^2}{(\sigma_L+\sigma_R)(\sigma_R-\sigma_L)}\right) \log\left(\frac{\sigma_R}{\sigma_L}\right)} \qquad (10)$$

If $\sigma_L$ and $\sigma_R$ are distributed around one value then $R_c$ is distributed at very high value. Based on figure 1, the situation will gives asymmetrical FYC. On contrary, if $\gamma$ is very high value then $R_c$ will be distributed at zero, hence this will gives symmetrical FYC. Figure 9 below shows these results,

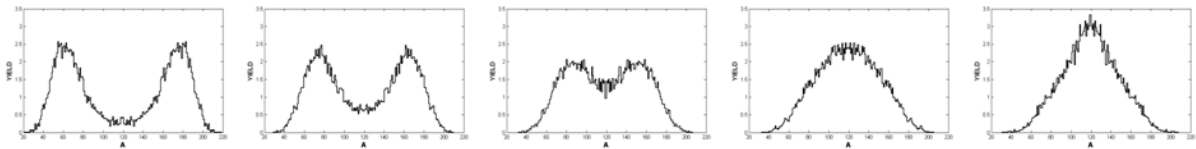

Figure 9. FYC results that they are calculated from various $\gamma$ values

Figure 9 demonstrates FYC of toy nuclide with A=238 results from various $\gamma$ values. From left to the right, $\gamma$ values are 0.1, 0.5, 1.0, 1.5, and 2.

Figure 10 shows comparison between FTM result and FTM modification result

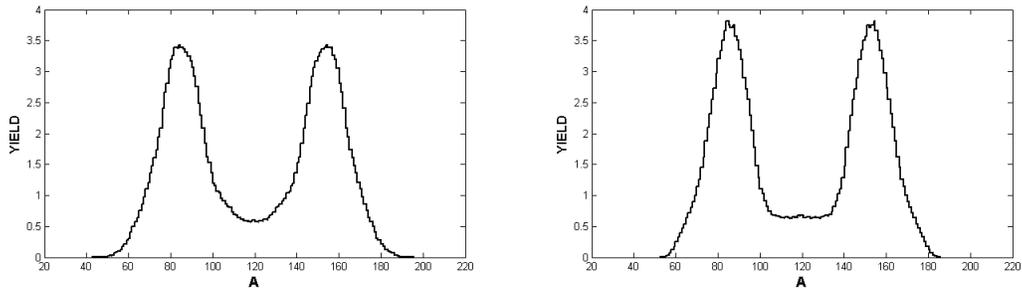

Figure 10. These are fission yields of 238-toy nuclide, left figure is taken from FTM and right figure is calculated from FTM modification.

Because of numerical approach, the FTM does not similar with FTM modification. The discrepancies are caused by interpolation of SDFD. In this step, there are errors form calculations of SDF. The errors are propagated through next steps; hence the results have discrepancies as shows by the figures. Even though there are some discrepancies, but the pattern of these figures are similar. The figures below show the influence of curve fitting in SDFD reproduction.

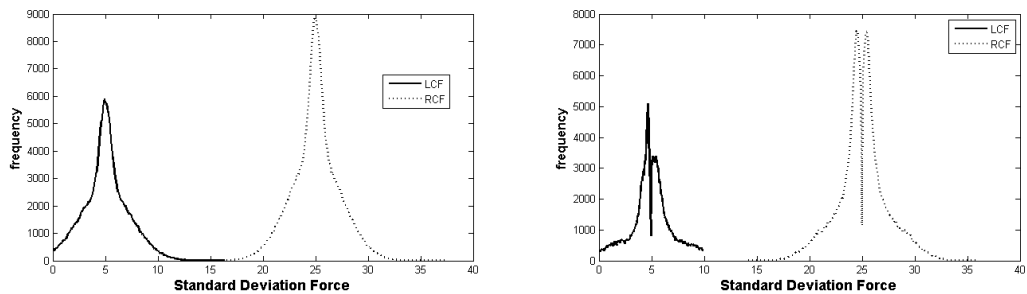

Figure 11. The SDFD of whole process, left figure is original SDFD,

right one is produced by FTM modification

Figure 11 show clearly that there are some errors from SDFD curve fitting step especially in center of curves. These discrepancies cause error propagation is multiplied; hence the fission yield is obtained as showed by figure 10.

**Conclusions**

The four figures (figure 4, 5, 6, and 7) have shown that $\sigma$ distribution strongly influences to fission yield curve. The influences of $\sigma$ distribution to fission yield curve can be reported as follows:

a. Higher $\mu_\sigma$ gives longer distance between two peak
b. Sharper of $\sigma$ distribution curve produces two kinds of fission yield curve, firstly if it has narrow curve at bottom frequency then fission yield is dominated by asymmetrical fission with two peaks. Secondly if it likes perturbation at blunt distribution then sharp distribution contributes at symmetrical fission process.

c.  If $\sigma_L$ distribution is significantly different from $\sigma_R$ then fission yield curve will have asymmetrical results.

Furthermore in order to obtain information comprehensively, mathematical formulation of LC and RC must be entangled.

There are two conclusions about variation of $\gamma$ values that are:

a.  $\gamma$ Value strongly influences to FYC, smaller $\gamma$ gives asymmetrical FYC and higher $\gamma$ gives symmetrical FYC.
b.  In real fission event, higher energy gives almost similar result with $\gamma=1.5$.

The FTM modification algorithm gives two conclusions:

a.  The algorithm can be implemented in calculation of fission yield if SDFD is available.
b.  The curve fitting must be avoided.

**Acknowledgement**

This research is fully supported by ITB Research Grant 2013
**References**

1. J.J. Cowan, F. –K. Thielemann, J. W. Truran, Phys. Rep. **208**, 267 (1991)
2. T. K. Kim, "*Gen-IV Reactors*", Nuclear Energy 2013, pp 175-201
3. Y. Kadi, A. H. Martinez, Nucl. Instrum. Meth. A **563**,574 (2006)
4. L. M. – Romao, D. Vendeplassche,"Accelerator Driven System", Conf. Proc. C12052016-10, 3rd Int. Part. Acc. Conf, Ne w Orleans, Louisiana, May 2012
5. G. H. Marcus, Prog. Nucl. Energy **37**, 5 (2000)
6. A. Aizawa, F. Kubo, T. Iwasaki, "Comparison of different neutronics analysis technique for Accelerator-Driven System", Annals of Nuclear Energy vol 60, October 2013, Pp 368-373
7. A. C. Wahl,"Compilation and evaluation of fission yields nuclear data",p 45-75, International Atomic Energy Agency, IAEA-TECDOC-1168 (2000)
8. H. Moriyama, T Ohnishi,"Systematic of fission fragment Mass-Yield Curves",Tech. Rep. Inst. Atom. Energy, Kyoto Univ., No 166 (1974)
9. J. Katakura,"A systematic of fission product mass yield with 5 gaussian functions", JAERI-Research, 2003-004.
10. K. Mazurek, C. Schmitt, P. N. Nadtochy, A. Maj, P. Wasiak, K. Kmiecik, B. Wasilewska,"Fission fragment mass distribution as a probe of the shape-dependent congruence energy term in the macroscopic models", Acta Physica Polonica B, 44, 293 (2013)
11. M. C. Duijvestijn, A. J. Koning, F. J. Hambsch,"Mass distribution in nucleon induced fission at intermediate energies",Physical Review C, Vol 64, 014607, (2001)
12. N. Schunck,"Microscopic description of induced fission",Journal of Physics" Conference Series 436 (2013) 012058

13. J. McDonnell, N. Schunck,"Microscopic Description of Nuclear Fission: Fission Barrier heights of Even-even Actinides", arXiv:1302.7587v1 [nucl-th] 2013
14. W. Younes, D. Gogny,"Fragment yields calculated in a time-dependent microscopic theory of fission", Technical report: Lawrence Livermore National Laboratory (LLNL), Livermore, CA, LLNL-TR-586678, 2012
15. H. Goutte, J. F. Berger, P. Casoli, and D. Gogny,"Microscopic approach of fission dynamics applied to fragment kinetic energy and mass distribution in 238U", Physical Review (2005)
16. U. Brosa, S. Grossman, A. Muller, "*Nuclear Scission*", Phys. Rep. 197, No. 4, pp 167-262, (1990)
17. C. Hagmann, J. Randrup, R. Vogt,"FREYA- A new Monte Carlo code for improved modeling of fission chains",Technical report: Lawrence Livermore National Laboratory (LLNL), Livermore, CA, LLNL-PROC-561431, 2012
18. D. Regnier, O. Litaize, O. Serot,"A Monte Carlo simulation of prompt gamma emission from fission fragments", EPJ Web of Conferences 42, 04003 (2013)
19. S. Hayakawa, M. Kawai, K. Kikuchi, "Nuclear reactions at moderate energies and Fermi gas model", Progress of Theoretical Physics, Vol. 13, No. 4, April 1955